\RequirePackage{fix-cm}

\documentclass[smallextended]{svjour3}       

\usepackage{amsmath}
\usepackage{amssymb}
\usepackage{array}
\usepackage{physics}
\usepackage[]{qcircuit}
\usepackage{graphicx}
\graphicspath{{./}}

\begin{document}

\title{The effect of quantum noise on algorithmic perfect quantum state transfer on NISQ processors}

\author{D.V. Babukhin$^{*}$        \and
        W.V. Pogosov 
}

\institute{D.V. Babukhin$^{*}$ \at
              Dukhov Research Institute of Automatics (VNIIA), 127055 Moscow, Russia \\
              Russian Quantum Center (RQC), 143026 Moscow, Russia \\\
              Tel.: +7(903)686-4804 \\\
              \email{dv.babukhin@gmail.com}           
           \and
           W.V. Pogosov \at
              Dukhov Research Institute of Automatics (VNIIA), 127055 Moscow, Russia \\
              Institute for Theoretical and Applied Electrodynamics, Russian Academy of
              Sciences, 125412 Moscow, Russia \\
              HSE University, 109028 Moscow, Russia\\\
              \email{walter.pogosov@gmail.com}           
}

\date{Received: date / Accepted: date}

\maketitle

\begin{abstract}
    Quantum walks are an analog of classical random walks in quantum systems. Quantum walks have smaller hitting times compared to classical random walks on certain types of graphs, leading to a quantum advantage of quantum-walks-based algorithms. 
	An important feature of quantum walks is that they are accompanied by the excitation transfer from one site to another, and a moment of hitting the destination site is characterized by the maximum probability amplitude of observing the excitation on this site. It is therefore prospective to consider such problems as candidates for quantum advantage demonstration, since gate errors can smear out a peak in the transfer probability as a function of time, nevertheless leaving it distinguishable. We investigate the influence of quantum noise on hitting time and fidelity of a typical quantum walk problem - a perfect state transfer (PST) over a qubit chain. We simulate dynamics of a single excitation over the chain of qubits in the presence of typical noises of a quantum processor (homogeneous and inhomogeneous Pauli noise,  crosstalk noise, thermal relaxation, and dephasing noise). We find that Pauli noise mostly smears out a peak in the fidelity of excitation transfer, while crosstalks between qubits mostly affect the hitting time.
	Knowledge about these noise patterns allows us to propose an error mitigation procedure, which we use to refine the results of running the PST on a simulator of a noisy quantum processor.  
            

\end{abstract}

\section{Introduction}
\label{sec: intro}

Classical random walks have many applications in computer science. They play an important role in randomized algorithms such as volume estimation of a convex body \cite{convex volume}, approximation of a matrix permanent \cite{permanent}, SAT-problems \cite{SAT} and combinatorial sampling problems \cite{combinatorial sampling}.  Their quantum counterparts extend these applications to quantum information, allowing for exponential speed-up in some problems (e.g., exponential faster hitting times \cite{fahri1998,childs2002,childs2003,kempe2003}), or just polynomial speed-up (e.g., quantum walk search problems \cite{shenvi2003,childs search,ambainis1,ambainis2}). There are applications of quantum walks to universality of quantum computation \cite{universal1,universal2},  photosynthesis and energy transfer \cite{photosynt}, implementation of quantum versions of PageRank algorithm \cite{QPageRank,QPageRank2}, and many more practically interesting problems. A quest to implement quantum walks efficiently in the era of noisy quantum devices is a challenging ongoing research field, where we can see considerable progress being made in recent years, see, e.g., Ref. \cite{62qubitQwalks} dealing with quantum walks on the 62-qubit superconducting quantum processor.


A particular example of quantum walks is a perfect state transfer \cite{Bose} - an exact transfer of a quantum state from node $A$ to node $B$ in particular time $T$ when the fidelity of the state transfer in a moment $T$ is unity. The problem is interesting from theoretical and practical use because it is a crucial element of quantum information processing on emerging quantum devices. In previous decades, many works investigated physical requirements for solid-state systems to be able to realize the perfect state transfer for arbitrary quantum states and in the presence of finite temperature baths (see \cite{perfect transfer1,perfect transfer2,perfect transfer3,perfect transfer4}) 
In recent years, several groups implemented the PST protocol on currently available quantum processor architectures \cite{Exp1,Exp2,Exp3,Feldman}, thus testing the ability of these devices to process quantum-encoded information. There were further investigations of decoherence influence on the PST \cite{decoherence1,decoherence2}. A perfect quantum state transfer is a promising tool for quantum information science, which allows for the preparation of entangled initial states, signal amplification \cite{Kay} and constructing quantum gates on a dynamical graph \cite{dynamic gate prep}.

Implementation of the perfect state transfer on digital quantum processors has its peculiarities.
In this setting, one needs to implement the PST using basis quantum gates, available on the particular quantum processor. Because these gates are not error-free on contemporary processors, one cannot implement the state transfer perfectly. A cumulative effect of gate errors and physical errors of the device (e.g., crosstalk interaction between solid-state qubits and energy relaxation of qubits) can completely corrupt the quantum-encoded information on the way across the quantum register. Since state transfer is an indispensable subroutine of quantum information processing, it is crucial to know how the typical noise of a quantum device affects its main properties.

In this paper, we investigate the cumulative influence of typical kinds of quantum noise in superconducting quantum processors on a hitting time and fidelity of the perfect state transfer over a chain of qubits. We implement the PST with Trotterization of an evolution operator and obtain dependencies of the PST characteristics on a depth of the quantum circuit for several kinds of quantum noise: single and two-qubit gate Pauli noise (homogeneous and inhomogeneous), crosstalk noise, finite energy relaxation time $T_1$, dephasing time $T_2$,  and composition of these noises as an error model of a noisy intermediate-scale quantum (NISQ \cite{nisq}) processor. We find that Pauli noise mostly smears out a peak in the fidelity of excitation transfer, while crosstalks between qubits mostly affect hitting time. Furthermore, we find that as long as homogeneous and inhomogeneous gate error noise have equal intensity, there is no difference in their influence on the PST quality.
Based on these patterns, we propose an error mitigation procedure, which we use to refine the results of running the PST on a simulator of a noisy quantum processor.

\section{Perfect state transfer over a chain of qubits}
\label{sec: pst}

The problem of a perfect quantum state transfer over a chain was first investigated by Bose \cite{Bose}. Criteria for the perfect state transfer in spin systems were developed in \cite{perfect transfer1,perfect transfer2,perfect transfer3}. First, an initial state of the system must have an integer number of excitations. Second, coupling constants should be of the form $J_{i} = C\sqrt{i(\mathcal{N}_{qubits}-i)}$, where $C$ is a constant, $\mathcal{N}_{qubits}$ is a chain length and $i$ is a number of a particular node. These criteria allow for a mirror symmetry in the system and the possibility to inverse a quantum state of the system quantum state during some unitary evolution \cite{perfect transfer4}. This evolution can be implemented with a hamiltonian
\begin{equation}
    \label{eqn: hamiltonian}
    H = -\sum_{i=1}^{\mathcal{N}_{qubits}-1}J_{i}(\sigma_{i}^{X}\sigma_{i+1}^{X} + \sigma_{i}^{Y}\sigma_{i+1}^{Y}).
\end{equation}
To implement this evolution on a digital quantum processor, one needs to use a Trotter decomposition of the evolution operator, which in the simplest case has the following form:
\begin{equation}
    e^{-i\Delta t(H_{A} + H_{B})} \approx e^{-i\Delta t  H_{A}/2}e^{-i\Delta t H_{B}/2} + O(\Delta t ^{2}),
\end{equation}
where $[H_{A}, H_{B}] \neq 0$. The Trotter decomposition allows constructing an evolution operator from a basic set of gates available on a particular quantum processor. For the hamiltonian (\ref{eqn: hamiltonian}) a corresponding quantum circuit can be constructed with the use of only two $CNOT$ gates (see Fig. 1 and  \cite{smithmanybody})
\begin{figure}[ht]
	\[
	\Qcircuit @C=1.0em @R=1.0em
	{
		\lstick{Q_{0}}	& \qw & \gate{HS^{\dagger}H} & \ctrl{1} & \gate{R_{x}(Jt)} & \ctrl{1} & \gate{HSH} & \qw \\
		\lstick{Q_{1}} & \qw & \gate{HS^{\dagger}H} & \targ 
		& \gate{R_{z}(Jt)} & \targ & \gate{HSH} & \qw \\
	}
	\]
	\caption{A quantum circuit, implementing an XY hamiltonian interaction among two qubits.}
\end{figure}
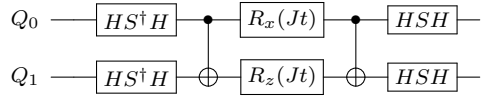
where $H$, $S$, $R_x$, $R_z$ and $CNOT$ gates are defined with matrices, provided in Fig. 2.

\begin{figure}[ht]
	\[
	\Qcircuit @C=1.0em @R=1.0em
	{
		\lstick{} & \gate{H} & \qw 
	}
	= 
	\frac{1}{\sqrt{2}}
	\begin{pmatrix}
	    1 & 1 \\ 1 & -1
	\end{pmatrix},
	\]
\end{figure}
\begin{figure}[ht]
	\[
	\Qcircuit @C=1.0em @R=1.0em
	{
		\lstick{} & \gate{S} & \qw 
	}
	= 
	\begin{pmatrix}
	    1 & 0 \\ 0 & i
	\end{pmatrix}.
	\]
\end{figure}
\begin{figure}[ht]
	\[
	\Qcircuit @C=1.0em @R=1.0em
	{
		\lstick{} & \gate{R_{x}(\theta)} & \qw 
	}
	= 
	\begin{pmatrix}
	    \cos{\theta/2} & -i\sin{\theta/2} \\ -i\sin{\theta/2} & \cos{\theta/2}
	\end{pmatrix},
	\]
\end{figure}
\begin{figure}[h!]
	\[
	\Qcircuit @C=1.0em @R=1.0em
	{
		\lstick{} & \gate{R_{z}(\theta)} & \qw 
	}
	= 
	\begin{pmatrix}
	    e^{-i\theta/2} & 0 \\ 0 & e^{i\theta/2}
	\end{pmatrix},
	\]
\end{figure}
\begin{figure}[h!]
	\[
	\Qcircuit @!
	{
		\lstick{} & \ctrl{1} & \qw \\
		\lstick{} & \targ & \qw 
	}
	= 
	\begin{pmatrix}
	    1 & 0 & 0 & 0 \\ 
	    0 & 1 & 0 & 0 \\ 
	    0 & 0 & 0 & 1 \\ 
	    0 & 0 & 1 & 0
	\end{pmatrix}.
	\]
	\caption{Typical quantum gates and their corresponding matrix representations}
\end{figure}

To quantify the efficiency of the state transfer over the qubit chain, one usually uses two quantities -  fidelity of the state transfer and a hitting time of the state transfer. Fidelity of the state transfer between qubits (nodes) $A$ and $B$ is defined as 
\begin{equation}
    \label{eqn: fidelity}
	F(t) = \bra{B} U(t) \ket{A},
\end{equation}
where $\ket{A}$ and $\ket{B}$ are states with one excitation residing on the initial qubit $A$ and the final qubit $B$, respectively. The state transfer is perfect if $F(t) = 1$.
A hitting time of the PST is a moment of time, when the PST fidelity is equal to unity 
\begin{equation}
    t = T_{PST}: F(T_{PST}) = 1.
\end{equation}
In the case of a non-perfect state transfer (e.g., due to errors during the evolution of quantum register), the hitting time is defined as a time, when the probability to find an excitation on the qubit $B$ is maximum over the evolution time:
\begin{equation}
    \label{eqn: hitting time}
    t = T_{hitting}: F(T_{hitting}) = \max\limits_{t}[ F(t)].
\end{equation}    
    
When running the PST on the NISQ device, quantum noise will affect the efficiency of the PST. In the following sections, we will discuss the main models of noise, typical to contemporary quantum processors, as well as opportunities for correcting the noise influence.

\section{Noise models}

\label{sec: noise models}

\subsection{Gate error noise}
Depolarizing noise is a homogeneous depolarization of a qubit state, which results in mixing of an initial state with a completely depolarized state. For a single qubit this noise is represented by a quantum channel
\begin{equation}
    \Phi^{depol}_{1q}(\rho) = (1 - q)\rho + \frac{q}{2}I,
\end{equation}
and is usually used as a benchmark model for errors of quantum gates. For two-qubit gates the quantum channel can be constructed out of single-qubit quantum channels in a following way 
\begin{equation}
    \Phi^{depol}_{2q}(\rho) = (\Phi^{depol}_{1q} \otimes \Phi^{depol}_{1q})(\rho).
\end{equation}
This construction assumes uncorrelated errors on different qubits.

A Pauli noise model is an extention of depolarizing model, when there is a preferred direction of depolarization in terms of axis of the Bloch sphere of a qubit. For a single qubit this noise is represented by a quantum channel
\begin{equation}
    \Phi^{Pauli}_{1q}(\rho) = (1 - p)\rho + p_{x}X\rho X + p_{y}Y\rho Y + p_{z}Z\rho Z,
\end{equation}
where $p$ is a probability of no error during the single-qubit gate, $p_{x}$, $p_{y}$ and $p_{z}$ are probabilities of occuring an $X$, $Y$, or $Z$ errors, respectively, i.e., application of an additional Pauli gate to the initial quantum gate. For two-qubit gates, assuming uncorrelated errors on different qubits, the two-qubit quantum channel is 
\begin{equation}
    \Phi^{Pauli}_{2q}(\rho) = (\Phi^{Pauli}_{1q} \otimes \Phi^{Pauli}_{1q})(\rho).
\end{equation}
Consider a particular case, when $X$, $Y$ and $Z$ errors are of equal probability. Then, we can use a following relation
\begin{equation}
    \frac{1}{2}I = \frac{\rho + X\rho X + Y\rho Y + Z\rho Z}{4}
\end{equation}
to relate parameters of the depolarizing and the Pauli noise models:
\begin{eqnarray}
    \notag
    \Phi^{depol}(\rho) = (1-q)\rho + q\frac{I}{2} = \\
    \notag
    (1 - \frac{3q}{4})\rho + \frac{q}{4}(X\rho X + Y\rho Y + Z\rho Z) = \\
    (1 - p)\rho + \frac{p}{3}(X\rho X + Y\rho Y + Z\rho Z).
\end{eqnarray}
We obtain a following relation of depolarizing and Pauli gate error noise intensities
\begin{equation}
    \label{eqn: intensities}
    q = \frac{4}{3}p,
\end{equation}
which we will use in the following.

\subsection{Crosstalks noise}
A crosstalk noise is an influence of the spurious interactions between solid-state qubits on dynamics of each other. This kind of interaction is ubiquitous in superconducting quantum processors \cite{crosstalk1,crosstalk3,crosstalk4}.
A ZZ crosstalk noise model, relevant for fixed-frequency qubits, describes a pairwise ZZ interaction between neighboring qubits - qubits, which can be subject to a two-qubit operation. It is represented by the two-qubit quantum channel
\begin{equation}
	\label{eqn:crosstalk dm}
    \Phi(\rho) = U_{ZZ}\rho U_{ZZ}^{\dagger},
\end{equation}
where
\begin{equation}
    U_{ZZ}(t) = 
    \begin{pmatrix}
        e^{-i\zeta t} & 0 & 0 & 0 \\
        0 & e^{i\zeta t} & 0 & 0 \\
        0 & 0 & e^{i\zeta t} & 0 \\
        0 & 0 & 0 & e^{-i\zeta t}
    \end{pmatrix},
\end{equation}
where $\zeta$ is a ZZ interaction constant. A presence of crosstalks affects circuits of intermediate and large depth, because its influence becomes stronger as longer physical time it takes to execute the circuit. Hereafter we assume a linear arrangement of qubits, so that ZZ crosstalks act  between neighboring qubits only. Nowadays quantum processors have crosstalk values of the order of $0.01 - 0.1$ MHz. The time duration of a two-qubit gate in this architecture is nearly 400 ns, so that the typical error due to the crosstalk on this scale is of the order of 1 percent. 

\subsection{Energy relaxation and dephasing noise}
Due to uncompensated coupling to external degrees of freedom, qubits are subject to energy relaxation and dephasing. These processes can also be described with quantum channels. For a single-qubit thermal relaxation process, a following quantum channel is used
\begin{equation}
    \Phi^{1q}_{T_1}(\rho) = 
    \begin{pmatrix}
    \rho_{00} - e^{-\frac{t}{T_1}}\rho_{11} & e^{-\frac{t}{2T_1}}\rho_{01} \\ 
    e^{-\frac{t}{2T_1}}\rho_{10} & e^{-\frac{t}{T_1}}\rho_{11}
    \end{pmatrix}.
\end{equation}
This quantum channel is parametrized by $T_1$ parameter, which characterizes a lifetime of qubit excitation. This channel can be written in terms of Kraus operators as \cite{Nielsen}
\begin{equation}
    \Phi^{1q}_{T_1}(\rho) = E_{1}\rho E_{1}^{\dagger} + E_{2}\rho E_{2}^{\dagger},
\end{equation}
where
\begin{equation}
    E_{1}
    =
    \begin{pmatrix}
    1 & 0 \\ 0 & \sqrt{1 - \gamma_{1}}
    \end{pmatrix},
\end{equation}
\begin{equation}
    E_{2}
    =
    \begin{pmatrix}
    0 & \sqrt{\gamma_{1}} \\ 0 & 0
    \end{pmatrix},
\end{equation}
and $\gamma_{1} = 1 - e^{-t/T_{1}}$. We rewrite $E_{1}$ in a way, which allows distilling the error-free part of the map:
\begin{equation}
    E_{1} = \sqrt{1 - \gamma_{1}}
    \begin{pmatrix}
    1 & 0 \\ 0 & 1
    \end{pmatrix}
    +
    (1 - \sqrt{1 - \gamma_{1}})
    \begin{pmatrix}
    1 & 0 \\ 0 & 0
    \end{pmatrix}.
\end{equation}
Thus, one can write a following form of the energy relaxation process channel
\begin{equation}
    \Phi^{1q}_{T_1}(\rho) = (1 - \gamma_{1})\rho + \Tilde{\rho},
\end{equation}
where
\begin{eqnarray}
    \notag
    \Tilde{\rho} = (1 - \sqrt{e^{-\frac{t}{T_1} }})\ket{0}\bra{0} \rho \ket{0}\bra{0} +     E_{1} \rho E^{\dagger}_{1} +
    \\
    2\sqrt{e^{-\frac{t}{T_1}}}(1 - \sqrt{e^{-\frac{t}{T_1}}})(\rho \ket{0}\bra{0} + \ket{0}\bra{0} \rho).
\end{eqnarray}
We can see that the error-free part of the quantum channel resulting state depends on the parameter of the channel in a way similar to a quantum channel of gate error noise. We will use this fact in the following section dealing with error mitigation.

For a single-qubit dephasing process, a following quantum channel is used
\begin{equation}
    \Phi^{1q}_{T_2}(\rho) = 
    \begin{pmatrix}
    \rho_{00} & e^{-\frac{t}{T_2}}\rho_{01} \\ 
    e^{-\frac{t}{T_2}}\rho_{10} & \rho_{11}
    \end{pmatrix},
\end{equation}
where $T_2$ characterizes qubit coherency lifetime. This channel can be written in terms of Kraus operators as \cite{Nielsen}
\begin{equation}
    \Phi^{1q}_{T_2}(\rho) = E_{1}\rho E^{\dagger}_{1} + E_{2}\rho E^{\dagger}_{2},
\end{equation}
where
\begin{equation}
    E_{1} 
    = 
    \sqrt{\gamma_{2}}
    \begin{pmatrix}
    1 & 0 \\ 0 & 1
    \end{pmatrix},
\end{equation}
\begin{equation}
    E_{2} 
    = 
    \sqrt{1 - \gamma_2}
    \begin{pmatrix}
    1 & 0 \\ 0 & -1
    \end{pmatrix}.
\end{equation}
Thus, one can write a following form of the dephasing process channel
\begin{equation}
    \Phi^{1q}_{T_2}(\rho) = \gamma_2\rho + (1 - \gamma_2)Z\rho Z .
\end{equation}
We again see, that the error-free part of the quantum channel resulting state depends on the parameter of the channel in a way, similar to gate error noise quantum channels.

To characterize energy relaxation and dephasing during worktime of two-qubit gates, one can use a tensor-product of single-qubit channels
\begin{equation}
    \Phi^{2q}_{T_1}(\rho) = (\Phi^{1q}_{T_1} \otimes \Phi^{1q}_{T_1})(\rho),
\end{equation}
\begin{equation}
    \Phi^{2q}_{T_2}(\rho) = (\Phi^{1q}_{T_2} \otimes \Phi^{1q}_{T_2})(\rho).
\end{equation}
If a single-qubit noise channel has a form, similar to channels described above, e.g.
\begin{equation}
    \Phi^{1q}(\rho) = (1 - P_{error})\rho + \Tilde{\rho},
\end{equation}
then, it transforms in the form of the two-qubit noise channel as 
\begin{equation}
    \Phi^{2q}(\rho) = (1 - P_{error})^{2}\rho + \tilde{\rho}.
\end{equation}
This pattern allows us to derive an error mitigation procedure which we will discuss in Sec. 4.

\section{Error mitigation}
\label{sec: error mitigation}

If dynamics are subject to noise, it will influence values of fidelity and hitting time. The exact model of cumulative noise of a quantum device is usually non-available in the process of computation. Methods like quantum process tomography \cite{Tomo1,Tomo2,Tomo3}, which in theory are capable of finding the exact model of noise, require an exponential amount of operations to be run on a quantum device. Furthermore, the reconstruction of noise quantum channels from raw data requires an exponential amount of resources for data processing. The demand for quantum process tomography makes it an impractical tool for characterizing the noise of a quantum device as soon as the device includes more than a dozen qubits.

Another route of working with noisy devices is to derive the main properties of noise from experimentally available data. We obtain knowledge about how cumulative noise affects the outcome of the computation through measuring observable values. If we also have some a-priori information about the noise structure, we can construct an error mitigation strategy based on experimentally available data. At the cost of being no more exact, this error mitigation strategy will improve our result with a resource cost, which has a crucial advantage against the full-knowledge-based method (quantum process tomography).

Let us apply this logic to the problem we study here. Consider we measure an observable $M$ to estimate the quality of state transfer over the system.  This observable acts on the Hilbert space of the target qubits, where we expect to find a transferred state after a particular time of evolution. If the final state of target qubits after time $t$ is $\rho(t)$, then we obtain 
\begin{equation}
    M(t) = Tr[M \rho(t)]
\end{equation}
If we use arguments of Sec.3, that we can distill an error-free part from every quantum noise channel among the standard set of channels (Markovian channels), then one can write a model-time dependence of the observable in the form
\begin{equation}
    M_{noisy}(t) = A(N)M_{ideal}(t + T_{shift}(N)) + B(N)
\end{equation}
where $A(N)$ and $B(N)$ are weights of error-free and error components of observable measurement result respectively, $T_{shift}(N)$ is a shift in model time, induced by quantum noise, $N$ is a circuit depth of the quantum circuit. Typical noise channels we consider here do not depend on parameters of gates they follow. Thus $A(N)$, $B(N)$ and $T_{shift}(N)$ do not depend on model time $t$ and only depend on depth of the quantum circuit $N$.

Using this model, we can construct an error mitigation procedure that allows the correction of both noise-induced rescaling and model-time shift. In the following, we consider a particular example of construction and use of the described approach. In particular, we will consider a case of the PST of a single-qubit excitation over a qubit chain. In this case, we can make further assumptions about $A(N)$, $B(N)$, and $T_{shift}(N)$ as functions of $N$. Knowing these forms, we propose methods for post-processing of results, serving as error mitigation.   

\subsection{Rescaling of dynamics}

In the noise models we consider here are two kinds of quantum channels - channels, which bring a qubit state towards maximally mixed state $\frac{1}{2}I$ and a thermal relaxation channel, which brings the state towards a qubit ground state $\ket{0}$. The composite noise channel thus must have an intermediate convergence point. In terms of the PST fidelity, thermal relaxation brings the fidelity value towards $Tr[\ket{0}\bra{0}\ket{1}\bra{1}] = 0$ and gate error and dephasing channels bring the fidelity value towards $Tr[\frac{1}{2}I \ket{1}\bra{1}] = \frac{1}{2}$. We can denote the state, corresponding to the composite channel stationary point as $\rho_{\alpha}$, so this channel brings the fidelity value towards $Tr[\rho_{\alpha}\ket{1}\bra{1}] = \alpha$. Thus, we can write down 
\begin{eqnarray}
    F = A(N) * 1 + \alpha(1 - A(N)).
\end{eqnarray}
As all quantum noise models introduce an exponential decay of the error-free part of the state (as provided in Sec.3), we assume $A(N) = c_{1}^{N}$. 

Under these assumptions, we fit fidelity dependencies,  obtained from simulations, using described noise models. These are calculated as $n_{noisy}(t) = |\bra{1}_{last}\tilde{n}\ket{\Psi(t)}|^{2}$, where $\Psi(t)$ is a state of the system in a moment $t$, $\bra{1}_{last} = \bra{0...01}$ is a basis state with a last qubit in an excited state and $\tilde{n} = (\frac{I - Z}{2})^{\otimes \mathcal{N}_{qubits}}$ is an $\mathcal{N}_{qubits}$-qubit excitation operator.
A fit function is defined as follows
\begin{equation}
    F = (1 - \alpha)c_{1}^{N} + \alpha,
\end{equation}
where $\alpha$ is a stationary value of fidelity. We can find this with time averaging of the last qubit excitation dependence, calculated with the maximal circuit depth:
\begin{equation}
    \alpha = \frac{1}{N_{T}}\sum_{t}n(t),
\end{equation}
where $N_{T}$ is the number of points along $t$ axis. To compensate for the contraction of observable dynamics, which arises from gate errors' mixing of quantum state, we need to re-scale data points of excitation dynamics:
\begin{equation}
    \hat{n}(t) = \frac{n(t) - \alpha(1 - c_{1}^N)}{c_{1}^N},
\end{equation}
where $c_{1}$ and $\alpha$ obtained from experimental data with fitting procedure proposed above.

\subsection{Model time shift}


As stated at the beginning of Sec.4, cumulative quantum noise also affects the model time. The main source of this effect is additional evolution, induced with crosstalk interaction between qubits. Previous works have illustrated how this evolution leads to oscillation of superposition state and that it can be strong enough to simulate an Ising-type dynamics \cite{TFIcrosstalks}. For this reason, it is indispensable to take this error into account.

In our simulations we worked in a regime where $|T_{hitting}(N) - T^{ideal}_{hitting}|$ is small compared to $T^{ideal}_{hitting}$. Thus, we used a linear approximation of time-shift function:
\begin{equation}
    T_{hitting}(N) = T^{ideal}_{hitting} + c_{2}N,
\end{equation}
which corresponds to the first linear term of series expansion of $T_{hitting}$ as a function of $N$. To compensate for this shift, we applied a shift of model time scale such that 
\begin{equation}
    t' = t - c_{2}N,
\end{equation}
where $N$ in the number of Trotter steps, used for a particular calculation and $c_2$ is obtained from fitting of experimental data.

\section{Results and discussion}

\subsection{Equality of homogeneous and inhomogeneous gate errors effect on the PST dynamics}

Here, we provide results of simulating the PST under the influence of homogeneous and inhomogeneous gate error noise. We demonstrate that it is sufficient to use a homogeneous gate errors model (depolarizing noise) to capture gate errors contribution in resulting dependencies. This observation validates our proposed error mitigation procedure.
 
We simulated the PST over a chain with the number of qubits $\mathcal{N}_{qubits} = 3$ for the following types of noise: 
\begin{enumerate}
    \item Homogeneous Pauli noise (depolarizing noise) with $q = 5*10^{-4}$
    \item Inhomogeneous Pauli noise with $p_x = \frac{3}{4}*3*10^{-4}$, $p_y = \frac{3}{4}*10^{-4}$, $p_z = \frac{3}{4}*10^{-4}$ and their cyclic permutations of $x$, $y$ and $z$.
    \item Pauli noise of a single type (X, Y or Z) with $p = 5*10^{-4}$
\end{enumerate}
We chose $\ket{\Psi_{0}} = \ket{1}$ as an initial state on the first qubit of the chain. The evolution operator is constructed as a Trotter decomposition of the evolution under the hamiltonian (\ref{eqn: hamiltonian}). A quantum circuit in Fig. 1 implements this evolution. We used coupling values $J_{12} = J_{23} = 2$ ($C = 2$) and observed dynamics on the time interval $[0.0, \pi]$ to capture a period of one Ising time $JT = 2\pi$. In these simulations, models of homogeneous and inhomogeneous gate error noise have equal intensities in a sense of (\ref{eqn: intensities}). 


We provide Table \ref{table: only pauli} and Table \ref{table: biased pauli} with model-time-averaged fidelity differences between depolarizing noise calculation and Pauli noise calculations. The fidelity difference value is defined as follows:
\begin{equation}
    \Delta^{fidelity}_{p_{x}, p_{y}, p_{z}} = \frac{1}{\mathcal{N}}\sum_{N}|F_{depol}(N) - F_{p_x, p_y, p_z}(N)|,
\end{equation}
where $p_x$, $p_y$ and $p_z$ are Pauli noise parameters and $\mathcal{N}$ is number of circuit depth setups: for example
we here considered $\mathcal{N} = 30$ circuit depth setups and $N = 1, 2, ..., 30$ number of Trotter steps. We see that homogeneous and inhomogeneous gate error noise with equal intensity (in a sense of (12)) make a similar effect on the PST. From Table \ref{table: only pauli} and Table \ref{table: biased pauli} we see that differences between dynamics with various Pauli noises and dynamics with depolarizing noise have similar values and severely differ from the fidelity difference between noise-free dynamics and dynamics with depolarizing noise. This allows us to consider depolarizing noise a good approximation of real gate error noise without loss of generality.

\begin{table}[ht]
 \begin{center}
 \begin{tabular}{| >{\centering\arraybackslash}m{1.5cm} | >{\centering\arraybackslash}m{1.5cm} | >{\centering\arraybackslash}m{1.5cm} | >{\centering\arraybackslash}m{1.5cm} | }
 \hline
  $\Delta^{fidelity}_{id}$ & $\Delta^{fidelity}_{0.05, 0, 0}$ & $\Delta^{fidelity}_{0, 0.05, 0}$ & $\Delta^{fidelity}_{0, 0, 0.05}$ \\
 \hline
    0.271 & 0.005 & 0.003 & 0.005 \\
 \hline
 \end{tabular}
 \end{center}
 \caption{Circuit-depth-averaged fidelity differences between excitation dynamics fidelity with depolarizing noise and with Pauli noise with a single type of Pauli error (X, Y or Z). A fidelity difference $\Delta^{fidelity}_{id}$ value for dynamics with depolarizing noise and noise-free dynamics is provided for reference.}
 \label{table: only pauli}
\end{table}
\begin{table}[ht]
 \begin{center}
 \begin{tabular}{| >{\centering\arraybackslash}m{2.0cm} | >{\centering\arraybackslash}m{2.0cm} | >{\centering\arraybackslash}m{2.0cm} | >{\centering\arraybackslash}m{2.0cm} | }
 \hline
  $\Delta^{fidelity}_{id}$ & $\Delta^{fidelity}_{0.03, 0.01, 0.01}$ & $\Delta^{fidelity}_{0.01, 0.03, 0.01}$ & $\Delta^{fidelity}_{0.01, 0.01, 0.03}$    \\ 
 \hline
  0.271 & 0.004
  & 0.003 & 0.004 \\
 \hline
 \end{tabular}
 \end{center}
 \caption{Circuit-depth-averaged fidelity differences between excitation dynamics fidelity with depolarizing noise and with biased Pauli noise. A fidelity difference $\Delta^{fidelity}_{id}$ value for dynamics with depolarizing noise and noise-free dynamics is provided for reference.}
 \label{table: biased pauli}
\end{table}
We also provide hitting time difference values defined as
\begin{equation}
    \Delta^{T_{hitting}}_{p_{x}, p_{y}, p_{z}} = \frac{1}{\mathcal{N}}\biggl|\sum_{N}T_{depol}(N) - T_{p_x, p_y, p_z}(N)\biggl|,
\end{equation}
which allow us to quantify how different noise types affect the PST hitting time. We present hitting time difference values for different gate error models in Table \ref{table: only pauli hitting time} and Table \ref{table: biased pauli hitting time}. We see, that there is a small difference between hitting time in the presence of depolarizing gate error and hitting time in the presence of different types of Pauli errors. We conclude, that using depolarizing gate error model in our simulations is justified.

\begin{table}[ht]
 \begin{center}
 \begin{tabular}{| >{\centering\arraybackslash}m{1.5cm} | >{\centering\arraybackslash}m{1.5cm} | >{\centering\arraybackslash}m{1.5cm} | >{\centering\arraybackslash}m{1.5cm} | }
 \hline
  $\Delta^{T_{hitting}}_{id}$ & $\Delta^{T_{hitting}}_{0.05, 0, 0}$ & $\Delta^{T_{hitting}}_{0, 0.05, 0}$ & $\Delta^{T_{hitting}}_{0, 0, 0.05}$ \\
 \hline
    0.036 & 0.002 & 0.003 & 0.009 \\
 \hline
 \end{tabular}
 \end{center}
 \caption{Circuit-depth-averaged hitting time differences between excitation dynamics hitting time with depolarizing noise and with Pauli noise with a single type of Pauli error (X, Y or Z). A hitting time difference $\Delta^{T_{hitting}}_{id}$ value for dynamics with depolarizing noise and noise-free dynamics is provided for reference.}
 \label{table: only pauli hitting time}
\end{table}

\begin{table}[ht]
 \begin{center}
 \begin{tabular}{| >{\centering\arraybackslash}m{2.0cm} | >{\centering\arraybackslash}m{2.0cm} | >{\centering\arraybackslash}m{2.0cm} | >{\centering\arraybackslash}m{2.0cm} | }
 \hline
  $\Delta^{T_{hitting}}_{id}$ & $\Delta^{T_{hitting}}_{0.03, 0.01, 0.01}$ & $\Delta^{T_{hitting}}_{0.01, 0.03, 0.01}$ & $\Delta^{T_{hitting}}_{0.01, 0.01, 0.03}$    \\ 
 \hline
  0.036 & 0.0
  & 0.001 & 0.012 \\
 \hline
 \end{tabular}
 \end{center}
 \caption{Circuit-depth-averaged hitting time differences between excitation dynamics hitting time with depolarizing noise and with biased Pauli noise. A hitting time difference $\Delta^{T_{hitting}}_{id}$ value for dynamics with depolarizing noise and noise-free dynamics is provided for reference.}
 \label{table: biased pauli hitting time}
\end{table}

\subsection{Post-correction of the PST dynamics}
In this section we discuss experiments, which demonstrate our idea about post-correction. We simulated the PST dynamics for the same system as in Sec. 5.1 for several cases:
\begin{enumerate}
    \item Dynamics without gate error noise, decoherence, and crosstalk noise
    \item Dynamics with gate error depolarizing noise and $T_1$, $T_2$ decoherence without crosstalk noise
    \item Dynamics with gate error depolarizing noise and $T_1$, $T_2$ decoherence with positive value crosstalk noise
    \item Dynamics with gate error depolarizing noise and $T_1$, $T_2$ decoherence with negative value crosstalk noise
\end{enumerate}

We simulated the perfect state transfer in the same system, described in Sec. 5.1.
We used parameters, inherent to state-of-the-art superconducting processors (e.g., see \cite{ibmqx}). In particular, we used $p_{1q} = 5*10^{-4}$ for single-qubit gate error noise and $p_{2q} = 1.28*10^{-2}$ for two-qubit gate error noise, $T_{1} = 80$ $\mu s$ and $T_{2} = 140$ $\mu s$, crosstalk values $\xi = \pm J/100 = \pm 0.01$. We used single qubit gate duration $L_{1q} = 35.5$ $ns$ and two-qubit gate duration $L_{2q} = 340$ $ns$ to simulate $T_{1}$ and $T_{2}$ decoherence.

\begin{figure}[ht]
	\centering
	\includegraphics[width=0.7\linewidth]{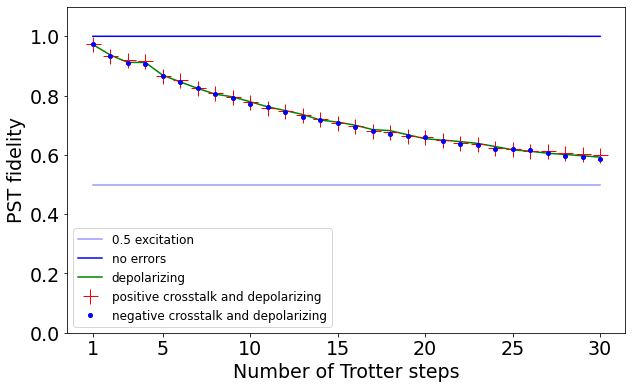}
	\caption{Fidelity dependencies on a circuit depth (measured in number of Trotter steps) for excitation dynamics in the presence of positive and negative crosstalks.}
	\label{fig:result fidelity plot}
\end{figure}
\begin{figure}[ht]
	\centering
	\includegraphics[width=0.7\linewidth]{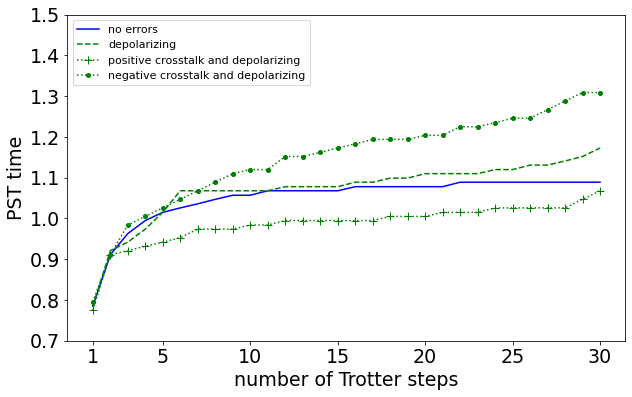}
	\caption{Hitting time dependencies on a circuit depth (measured in number of Trotter steps) for excitation dynamics in the presence of positive and negative crosstalks.}
	\label{fig:result hitting time plot}
\end{figure}
In Fig.\ref{fig:result fidelity plot} and Fig.\ref{fig:result hitting time plot}, we provide PST fidelity and hitting time dependencies on the circuit depth (measured in the number of Trotter steps) for four experiments mentioned above. For more than 5 Trotter steps, we can see that it is consistent with the pattern of gate error and crosstalk noise influence we described in Sec.3. We can see that crosstalk noise mostly affects hitting time while only slightly affecting fidelity. Then, adding depolarizing noise and incorporating decoherence processes associated with $T_1$ and $T_2$, we see exponential decay of the fidelity.

At low circuit depth ($N \leq 5$ Trotter steps) there is a root-like dependence of PST hitting time. We connect this behavior with an insufficient Trotter decomposition precision of the evolution operator. This behavior vanishes as the precision of Trotter decomposition saturates. We also note that unity fidelity is achieved even at small Trotter steps numbers despite the Trotterization errors. However, the hitting time, in this case, is not accurate, as seen from Fig.\ref{fig:result hitting time plot}.

\begin{table}[ht]
 \begin{center}
 \begin{tabular}{| >{\centering\arraybackslash}m{2.2cm} | >{\centering\arraybackslash}m{2.2cm} | >{\centering\arraybackslash}m{2.2cm} | >{\centering\arraybackslash}m{2.2cm} | }
 \hline
   noise-free & depolarizing & depolarizing and positive crosstalk & depolarizing and negative crosstalk    \\ 
 \hline
   1.0 & 0.954
   & 0.951 & 0.953 \\
 \hline
 \end{tabular}
 \end{center}
 \caption{$c_1$ fitting constants, calculated for simulation data for noiseless dynamics, dynamics with depolarizing gate error noise, dynamics with depolarizing gate error noise with positive and negative crosstalks.}
 \label{table: c1}
\end{table}

\begin{table}[ht]
 \begin{center}
 \begin{tabular}{| >{\centering\arraybackslash}m{2.2cm} | >{\centering\arraybackslash}m{2.2cm} | >{\centering\arraybackslash}m{2.2cm} | >{\centering\arraybackslash}m{2.2cm} | }
 \hline
   noise-free & depolarizing & depolarizing and positive crosstalk & depolarizing and negative crosstalk    \\ 
 \hline
   0.005 & 0.007
   & 0.005 & 0.013 \\
 \hline
 \end{tabular}
 \end{center}
 \caption{$c_2$ fitting constants, calculated for simulation data for noiseless dynamics, dynamics with depolarizing gate error noise, dynamics with depolarizing gate error noise with positive and negative crosstalks.}
 \label{table: c2}
\end{table}
In Table \ref{table: c1} we provide $c_{1}$ values from fitting fidelity dependencies in different experiments and in Table \ref{table: c2} we present $c_{2}$ values from fitting hitting time dependencies.
\begin{figure}[ht]
	\centering
	\includegraphics[width=0.7\linewidth]{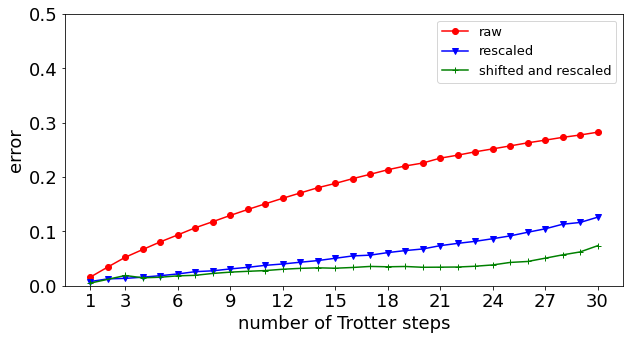}
	\caption{Excitation dynamics error for simulation data without postprocessing (circles), for data with rescaling (triangles) and for data with rescaling and model time shift (crosses)}
	\label{fig: metric crosstalk depol}
\end{figure}
In Fig.\ref{fig: metric crosstalk depol} we show excitation dynamics error before and after post-processing procedures for the case of depolarizing error with positive crosstalk. 
We define the excitation dynamics error as 
\begin{equation}
    error(N) = \sum_{t}|n_{0}(N,t) - n_{noisy}(N,t)|,
\end{equation}
where $n_{0}(N,t)$ and $n_{noisy}(N,t)$ are noise-free  and noisy data, respectively, for a circuit of $N$ Trotters steps.

\begin{figure}[ht]
	\centering
	\includegraphics[width=0.7\linewidth]{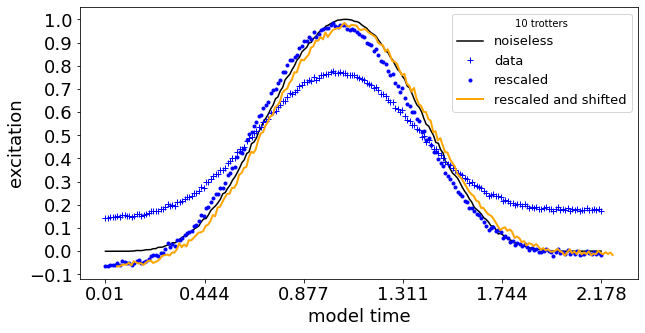}
	\caption{Excitation dynamics of the last qubit in the chain of 3 qubits under the influence of the depolarizing noise during the state transfer for 10 Trotter steps}
	\label{fig: depol exc 10}
\end{figure}
\begin{figure}[ht]
	\centering
	\includegraphics[width=0.7\linewidth]{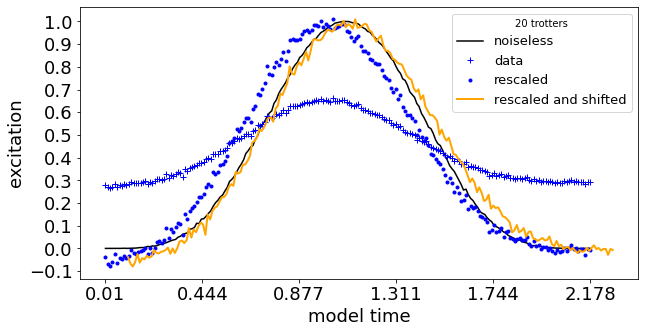}
	\caption{Excitation dynamics of the last qubit in the chain of 3 qubits under the influence of the depolarizing noise during the state transfer for 20 Trotter steps}
	\label{fig: depol exc 20}
\end{figure}

We show the excitation dynamics on the last qubit before and after rescaling and model time scale shift for $N = 10$ and $20$ in Figs. \ref{fig: depol exc 10} and \ref{fig: depol exc 20} correspondingly. We see that the shift of the hitting time grows as $N$ increases. We see that the described error mitigation techniques allow correction of a significant part of the total error and bring the dynamics much closer to the error-free case. 
In particular, rescaling almost perfectly allows us to restore the error-free range of excitation dynamics, and linear shift allows us to compensate for crosstalks' influence.

\section{Conclusions}
\label{sec: conclusion}

In this paper, we investigated the influence of two main sources of errors in quantum computing on the algorithmic transfer of a qubit excitation across the quantum register. We found that gate errors mostly affect the fidelity of the transfer, while ZZ residual interactions between qubits (crosstalks) mostly lead to a shift of the perfect transfer hitting time. Thus, gate errors smear the peak in the probability amplitude as a function of time but generally leave it distinguishable as long as circuit depth is not too large, while crosstalk's influence turns out to be more negative and leads to inaccurate hitting time determination. 

We demonstrated, that we only need to know the general properties of noise models without precise knowledge of noise parameters to mitigate its' effect on the computation results. In particular, we need to know that gate errors stack up to exponential decay of the error-free part of the final quantum state, while we do not need to know if these errors are homogeneous or not. 
Using our post-processing error mitigation, we were able to sufficiently improve the quality of results for circuits of about 270 layers depth. Our results provide an example of post-processing error mitigation of quantum computing outcomes without precise knowledge of the quantum noise model. 

Developing methods of quantum error mitigation is an important part of contemporary quantum computing science in the era of NISQ \cite{nisq} devices. Looking for real-world applications of these quantum devices was a topic of great interest in recent years. There are attempts to use currently available devices for solving problems in finance \cite{QFin1,QFin2,QFin3}, developing machine learning algorithms \cite{QML0,QML1,QML2} and solving quantum chemistry problems \cite{QChem1,QChem2,QChem3}. 
A promising path towards efficient error mitigation is using a toolbox of modern data science, e.g., applying machine learning and statistical analysis to classical post-processing of data (see, e.g., Refs. \cite{MLpostproc1,MLpostproc2}). 

\section*{Acknowledgements}
\label{sec: acknowledgements}

D. V. B. acknowledges a support from RFBR (project no. 20-37-70028)
and
W. V. P.
acknowledges a support from RFBR (project no. 19-02-00421).

\bibliographystyle{elsarticle-num} 

\end{document}